\documentclass[aip,apl,reprint,a4paper,floatfix,amsmath,amssymb,amsfonts,noshowpacs,citeautoscript]{revtex4-2}
\usepackage{newtxtext,newtxmath}
\usepackage[T1]{fontenc}
\usepackage{graphicx}
\usepackage[dvipsnames]{xcolor}
\usepackage{soul} 
\usepackage{floatrow}
\usepackage{siunitx}
\usepackage[
,textwidth=17.5cm
,textheight=23.5cm
,verbose
,pdftex
]{geometry}
\usepackage{xspace}
\usepackage{svg}
\usepackage[pdftex]{hyperref}
\hypersetup{
  pdftitle={Excitonic and deep-level emission from N- and Al-polar homoepitaxial AlN layers grown by molecular beam epitaxy},
	colorlinks,
	citecolor=blue,
	linkcolor=blue
	}

\makeatletter
\newcommand*{\refig}[2]{\hyperref[#1]{\ref*{#1}(#2)}}
\makeatother

\DeclareMathAlphabet{\mathsf}{OT1}{\sfdefault}{m}{n}
\SetMathAlphabet{\mathsf}{bold}{OT1}{\sfdefault}{b}{n}

\begin{document}

\preprint{AIP/123-QED}

\title[Sample title]{Edge emission from 265~nm UV-C LEDs grown by MBE on bulk AlN}
\author{Shivali Agrawal}
\email{sa2368@cornell.edu}
\affiliation{%
Department of Chemical and Biomolecular Engineering, Cornell University, Ithaca, USA
}
\author{Hsin-Wei S. Huang}
\email{hh494@cornell.edu}
\affiliation{Department of Electrical and Computer Engineering, Cornell University, Ithaca, USA}

\author{Debaditya Bhattacharya}
\affiliation{Department of Electrical and Computer Engineering, Cornell University, Ithaca, USA}

\author{Madhav Ramesh}
\affiliation{Department of Electrical and Computer Engineering, Cornell University, Ithaca, USA}

\author{Krzesimir Szkudlarek }
\affiliation{Institute of High Pressure Physics PAS (Unipress), Warsaw, Poland}

\author{Henryk Turski}
\affiliation{Department of Electrical and Computer Engineering, Cornell University, Ithaca, USA}
\affiliation{Institute of High Pressure Physics PAS (Unipress), Warsaw, Poland}

\author{Vladimir Protasenko}
\affiliation{Department of Electrical and Computer Engineering, Cornell University, Ithaca, USA}

\author{Huili Grace Xing}
\affiliation{Department of Electrical and Computer Engineering, Cornell University, Ithaca, USA}
\affiliation{Department of Materials Science and Engineering, Cornell University, Ithaca, USA}
\affiliation{Kavli Institute at Cornell for Nanoscale Science, Cornell University, Ithaca, USA}

\author{Debdeep Jena}
\email{djena@cornell.edu}
\affiliation{Department of Electrical and Computer Engineering, Cornell University, Ithaca, USA}
\affiliation{Department of Materials Science and Engineering, Cornell University, Ithaca, USA}
\affiliation{Kavli Institute at Cornell for Nanoscale Science, Cornell University, Ithaca, USA}
\begin{abstract}
UV-C LEDs pseudomorphically grown by MBE on bulk AlN substrates emitting at 265~nm are demonstrated. High current density up to 800~A/cm$^2$, 5 orders of on/off ratio, and low differential on-resistance of 2.6~m$\Omega\cdot$cm$^2$ at the highest current density is achieved. The LED heterostructure has a high refractive index waveguide core surrounded by n- and p-cladding layers similar to a laser diode designed for mode confinement at 270~nm to facilitate edge emission and collection of photons. Edge-emitting devices are made by cleaving the fabricated LEDs along the $m$-plane of the wurtzite crystal. Electrical injection results in emission of high energy 4.7~eV photons that are collected from the cleaved edge of the LEDs corresponding to the optical bandgap of the AlGaN active region. The contribution of power dissipation across the n- and p-regions of the diode is discussed. The n-contact resistance to n-AlGaN is identified as the largest contributor to the series resistance of the LED in the present generation of devices. 
\end{abstract}

\maketitle
High Al composition AlGaN-based UV-C (200 nm $<\lambda_\text{emission}<$ 280~nm) light emitting diodes (LEDs) and laser diodes (LDs) on bulk AlN substrates are attractive for the development of deep ultraviolet light sources. These diodes have applications in sterilization, disinfection, lithography, and machining\cite{kneissl2019emergence}. However these LEDs currently face challenges with their low external quantum efficiency, short lifetimes and rapid degradation at high injection\cite{kneissl2019emergence,zhang2024study}. Investigation of these structures using different growth techniques can provide valuable insights into solving the current challenges. Metal-organic chemical vapor deposition (MOCVD) and molecular beam epitaxy (MBE) remain the most common growth techniques for the development of these diodes. Recently reported UV-C laser diodes grown by MOCVD on bulk AlN substrates have shown impressive performance \cite{zhang2022key} but are still plagued by short lifetimes \cite{zhang2024study}.

While MOCVD is a scalable and commercially relevant growth technique, MBE offers ultra-high vacuum environment and hydrogen-free growth with ultrapure elemental sources which result in films with low C, H, Si, and O impurities as seen by time-of-flight secondary ion mass spectrometry (SIMS) measurements \cite{cho-2020}. Additionally, lower growth temperatures in MBE result in reduced passivating defects that tend to form in high temperature epitaxy\cite{chen2021gan}.

The crystal quality of the underlying AlN is important for device performance \cite{kobayashi2024enhanced}. AlGaN films grown on bulk AlN substrates with low threading dislocation density (TDD) increase the internal quantum efficiency (IQE), carrier injection efficiency (CIE), and lifetime of devices \cite{zhang2021polarization,bagheri2022doping,ruschel2020reliability,ren2007algan}.  An earlier study has demonstrated the first MBE grown optically-pumped 280~nm AlGaN-based laser on a bulk AlN substrate\cite{van2022optically}, highlighting the optical quality of AlGaN thin films grown by MBE. We also recently demonstrated the first MBE grown high Al content AlGaN p-n junction diodes appropriate for electrical injection on bulk AlN substrates \cite{agrawal2024ultrawide}. In this study, we demonstrate and characterize electrically injected edge emitting DUV LEDs enabled by MBE grown heterostructures on bulk AlN substrates.     
\begin{figure*}
    \includegraphics[width=\textwidth]{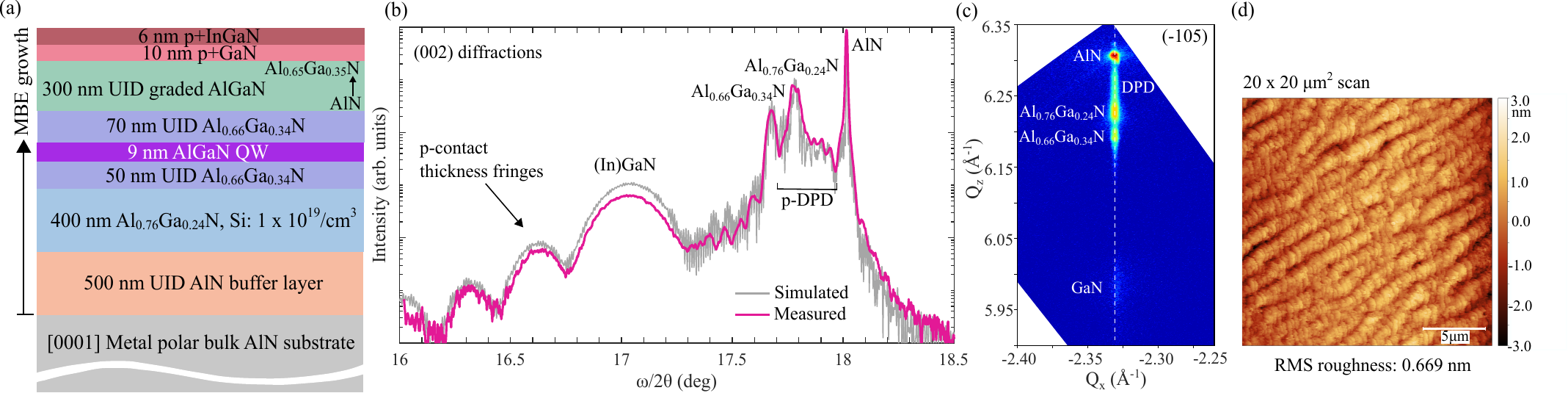}
    \caption{(a) Heterostructure of the LED used in this study. (b) Measured and simulated symmetric 2$\theta$-$\omega$ X-ray diffraction scans across the (002) planes. (c) Reciprocal space map across the asymmetric ($\bar{1}$05) diffractions. (d) 20 $\times$ 20 \textmu m$^2$ AFM scan on the surface of the LED.}
    \label{fig:Figure 1}
\end{figure*}

\begin{figure*}[t]
   \includegraphics[width=\textwidth]{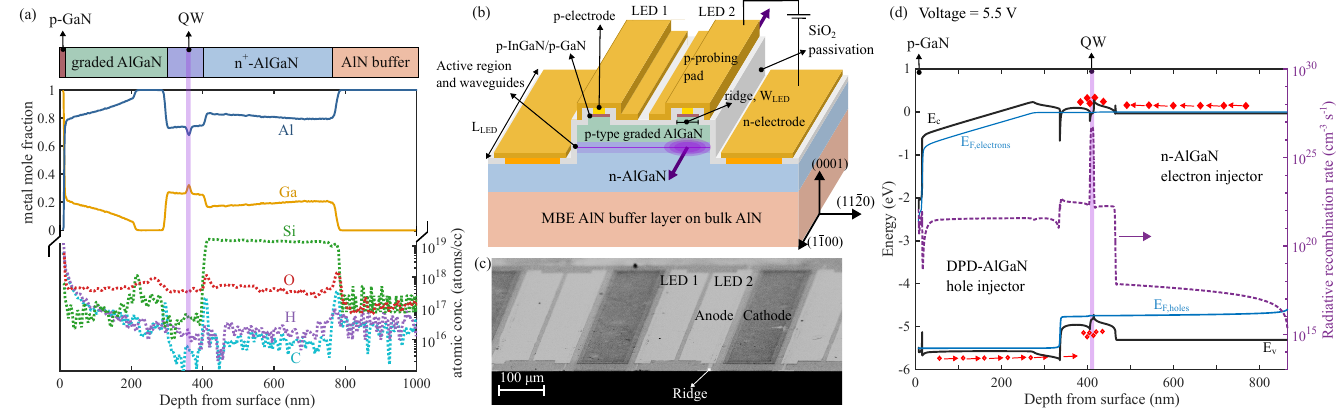}
    \caption{(a) Secondary ion mass spectrometry (SIMS) of the LED epilayers. (b) Device schematic of the fabricated LED structure. (c) SEM image of 200 \textmu m cavity length devices. (d) Energy band diagram and radiative recombination rate for the LED at 5.5~V forward bias.}
    \label{fig:Figure 2}
\end{figure*}

As the core of the LED is a pn junction, there is a need for electron and hole injection layers.  For hole injection in ultrawide-bandgap (UWBG) AlGaN, successful use of dopant-free distributed polarization doping (DPD) in high Al content AlGaN pn diodes has been realized in previous studies by both MBE and MOCVD\cite{agrawal2024ultrawide,ramesh2025ultrawide,kumabe2024demonstration}. As shown in Fig.~\ref{fig:Figure 1}(a), for electron injection we design a heterostructure with a heavily silicon doped n$^+$-AlGaN region. Combining the polarization and impurity doped layers for p- and n- regions, respectively, we demonstrate a step towards developing a UV-C laser. 

We implement a separate confinement heterostructure (SCH) diode with an undoped waveguide layer to confine photons, and a single quantum well to confine electron and hole carriers for recombination. The asymmetric waveguide structure is optimized to have the highest optical mode confinement factor $\Gamma$ of $\sim6\%$ at 270~nm. We fabricate Fabry-Pérot cavities with cleaved mirror facets. The emitted photons are waveguided in the high refractive index cladding layer and collected from the edge of the cavity.


The LED heterostructure shown in Fig.~\ref{fig:Figure 1}(a) was grown by nitrogen plasma-assisted MBE on a single crystal bulk AlN substrate. The targeted AlGaN epilayers on the metal polar face of the substrate along the $[0001]$ crystal direction are: a 500~nm AlN buffer layer, 400~nm n$^+$-Al$_{0.76}$Ga$_{0.24}$N with Si donor doping, a 9~nm single quantum well (SQW) sandwiched in a 120~nm asymmetric Al$_{0.66}$Ga$_{0.34}$N waveguide layer (50~nm on the n-side and 70~nm on the p-side), a 300~nm unintentionally doped (UID) AlGaN layer linearly graded from AlN to Al$_{0.65}$Ga$_{0.35}$N along the growth direction resulting in distributed polarization doping creating an effective $\rho_{\pi}\sim 6 \times 10^{17}$cm$^{-3}$ mobile hole density\cite{agrawal2024ultrawide}, and a Mg doped contact layer consisting of 10~nm p$^+$GaN and 6~nm p$^+$In$_{0.05}$Ga$_{0.95}$N. 

The targeted Al mole fraction was $\sim54\%$ in the quantum well for the desired 270 nm wavelength. The symmetric (002) X-ray diffraction scan is shown in Fig.~\ref{fig:Figure 1}(b) along with the theoretical diffraction pattern from a dynamical diffraction model. The extracted AlGaN and \mbox{InGaN} compositions are close to the targeted values. The clearly resolved interference fringes from the thin p-contact layer suggest sharp interfaces throughout the structure and are used to determine the p-contact layer thickness. The asymmetric reciprocal space map of the ($\bar{1}05$) X-ray diffraction scan shown in Fig.~\ref{fig:Figure 1}(c) suggests that all the AlGaN layers are strained to the AlN substrate as desired. Fig. \ref{fig:Figure 1}(d) shows the atomic force microscopy (AFM) image of the top p-InGaN surface after growth, exhibiting a smooth surface with sub-nanometer surface roughness over 20 $\times$ 20~\textmu m$^2$ scan area. 

\begin{figure*}
    \includegraphics[width=0.8\textwidth]{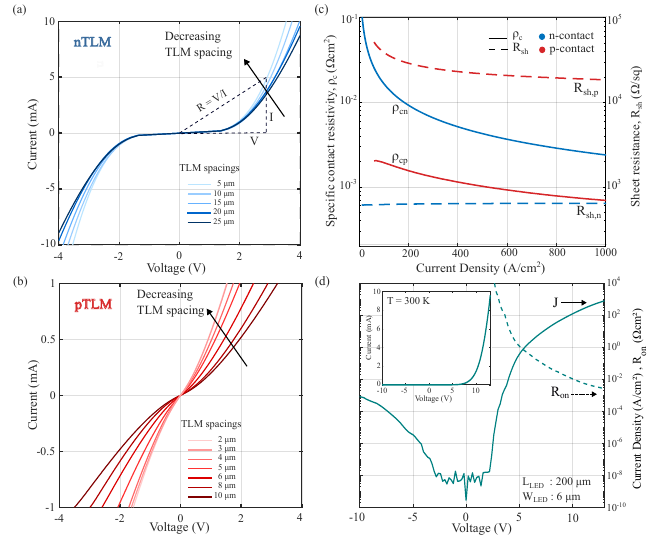}
    \caption{(a) TLM IVs of the n-contacts. The resistance is calculated at each current value using $R=V/I$ as shown in the figure. (b) TLM IVs of the p-contacts. (c) Specific contact resistivity and sheet resistance of the n- and p-contacts. (d) Linear (inset) and logarithmic diode IV characteristics and specific on resistance ($R_{\text{on}}$) at room temperature.}
    \label{fig:Figure 3}
\end{figure*}

SIMS depth profiles performed at EAG laboratories for Al, Ga, and impurity atoms C, H, Si, and O results are plotted in Fig.~\ref{fig:Figure 2}(a). The Al and Ga metal mole fraction profiles confirm the growth of high Al content Al$_x$Ga$_{1-x}$N throughout the structure desired for UV-C LEDs and lasers. The relative comparison of the AlGaN composition in the different epilayers is consistent between XRD and SIMS. 
Expected features like the presence of a quantum well and intentional linear composition grading of the AlGaN distributed polarization doped (DPD) layer are confirmed. 

We observe a $\sim$70~nm unintentional AlN layer formation at the Al$_{0.66}$Ga$_{0.34}$N p-waveguide and the p-DPD interface. Since Al has a strong thermodynamic preference for incorporation over Ga in metal rich MBE growth \cite{hoke2007thermodynamic}, the Al/N ratio is varied to control the AlGaN composition. Hence the formation of the AlN interlayer is likely due to an overshoot of the Al metal flux such that $\phi_\text{Al}/\phi_\text{N}>1$ causing the growth of binary AlN layer near the interface. Although this UID AlN serves as a desirable electron-blocking layer (EBL), it also is a barrier for hole injection and increases the diode on resistance. The SIMS analysis in the n-AlGaN layer confirms the Si doping level of ~$\sim 10^{19}$~cm$^{-3}$. The C and H levels are at the detection limits throughout the sample. The O level in MBE grown AlN is at the detection limit but is higher in AlGaN layers due to the lower growth temperature of AlGaN than AlN.

Quasi-vertical LEDs were fabricated from the epilayers as shown in Fig.~\ref{fig:Figure 2}(b). Device mesas were formed using chlorine based inductively coupled plasma reactive ion etching (ICP-RIE) $\sim$100~nm into the n-AlGaN layer. An additional ridge etch of $\sim$200~nm was performed into the DPD layer to define the current aperture. A V/Al/Ni/Au $(20/80/40/100~\text{nm}$) metal stack was deposited on the etched n-AlGaN surface using electron beam evaporation and annealed at 800$^{\circ}$C for 30~s in N$_2$ ambient. A Ni/Au ($15/20$~nm) metal stack was deposited on the p-InGaN and annealed at $450^{\circ}$C for 30~s in O$_2$ ambient and followed by deposition of Ti/Au ($20/100$~nm) probe pads. SiO$_2$ was deposited by plasma-enhanced atomic layer deposition (PE-ALD) for 50~nm and then by \mbox{plasma-enhanced chemical vapor deposition} (PECVD) for 200~nm to passivate the device surfaces. 

After fabrication into diodes, the AlN substrate was thinned down from $\sim$550~µm to $\sim$100~µm using mechanical polishing and then cleaved along the (1$\bar{1}$00) $m$-plane of the crystal into 200~µm to 1000~µm lengths to define Fabry-Pérot cavities. Fig.~\ref{fig:Figure 2}(c) shows the scanning electron microscope (SEM) image of cleaved cavity LED devices with a length \mbox{$L_{\text{LED}}=$ 200~µm} and p-electrode widths that range from \mbox{$W_\text{LED}=$ 2-20~µm}. The anode probing pad, cathode, and the ridge are labeled on the SEM image. The energy band diagram for the LED heterostructure confirmed by SIMS in Fig.~\ref{fig:Figure 2}~(a) is simulated using SiLENSe\cite{SiLENSe} at 5.5~V forward bias and is shown in Fig.~\ref{fig:Figure 2}(d). Radiative recombination is expected to occur in the QW region by design, as shown by the simulated radiative recombination rates at 5.5~V.

Transfer length method (TLM) pads in rectangular geometry were used to evaluate specific contact resistances of both n- and p- type semiconductor-metal contacts. As shown in Fig.~\ref{fig:Figure 3}(a), the current-voltage (IV) relationship between TLM pairs of n-contact shows barrier-limited transport characteristics and the total resistance shows proportional relationship to TLM pair spacing. To evaluate the non-linear IVs, apparent contact resistance $V/I$ was evaluated at constant current levels for various TLM spacings. The current density is calculated using the effective contact area which is calculated as TLM pad-width $W_{\text{TLM}}$~$\times$~$L_\text{t}$ (transfer length). Due to the non-linearity of IVs, the specific contact resistivity $\rho$$_{\mathrm{c}}$ shows dependence on current density $J$ as shown in Fig.~\ref{fig:Figure 3}(c). We acknowledge the uncertainty in the definition of $L_{\text{t}}$ in the case of non-linear contacts.  Nonetheless, using a current dependent $L_{\text{t}}$ allows the following rough analysis. The n-contact specific contact resistivity varies from 28~m$\Omega\cdot$cm$^2$ at $J\sim50$~A/cm$^2$ to 4.3~m$\Omega\cdot\mathrm{cm}^2$ at a higher current density of $J\sim500$~A/cm$^2$. The extracted sheet resistance of n-AlGaN is $\sim640~\Omega/\square$.  From an estimate of mobility $\mu = 30$ cm$^2$/V$\cdot$s, and mobile electron concentration $n=10^{19}$cm$^{-3}$ assuming full ionization of Si donors, and n-AlGaN layer thickness $t$ after etching = 300~nm, the sheet resistance $R_\mathrm{sh}=\frac{1}{q\cdot\mu\cdot n\cdot t}\sim690~\Omega/\square$, which is slightly higher than the $R_\mathrm{sh}$ value extracted from TLM. This suggests high ionization of Si dopants in the n-AlGaN layer and high mobility of electrons. The barrier-limited transport behaviour is likely due to a Schottky barrier and damage caused by the ICP-RIE process and needs to be further studied.

The p-contacts yielded a slightly lower specific contact resistivity of $\sim2$~m$\Omega\cdot$cm$^2$ at low current density of $J\sim50$~A/cm$^2$ and $\sim0.9$~m$\Omega\cdot$cm$^2$ at $J\sim500$~A/cm$^2$ as shown in Fig.~\ref{fig:Figure 3}(c). The extracted sheet resistance of p-contact layer is $\sim25$~k$\Omega/\square$. This is similar to the p-contact conductivity determined by Hall-effect measurement prior to fabrication: $6.8\times10^{13}~$cm$^{-2}$ sheet density, 2.71~cm$^2/$V$\cdot$s mobility and 33.9~k$\Omega/\square$ sheet resistance. The average mobile hole density in the p-contact layers is therefore \mbox{$6.8\times10^{13}~$cm$^{-2}/16 \times 10^{-7}$cm} = $4.25 ~\times$~10$^{19}$cm$^{-3}$. The high hole concentration and low p-contact resistivity were possible due to the thin InGaN cap layer\cite{9046469}. Several factors leading to lower specific contact resistivity include: higher activation of Mg dopants in InGaN compared to GaN, favorable 2D hole gas in InGaN induced by polarization charge at the GaN/InGaN heterojunction, and lower Schottky barrier height leading to better tunneling.

\begin{figure}[t]
   \includegraphics[width=8.5 cm]{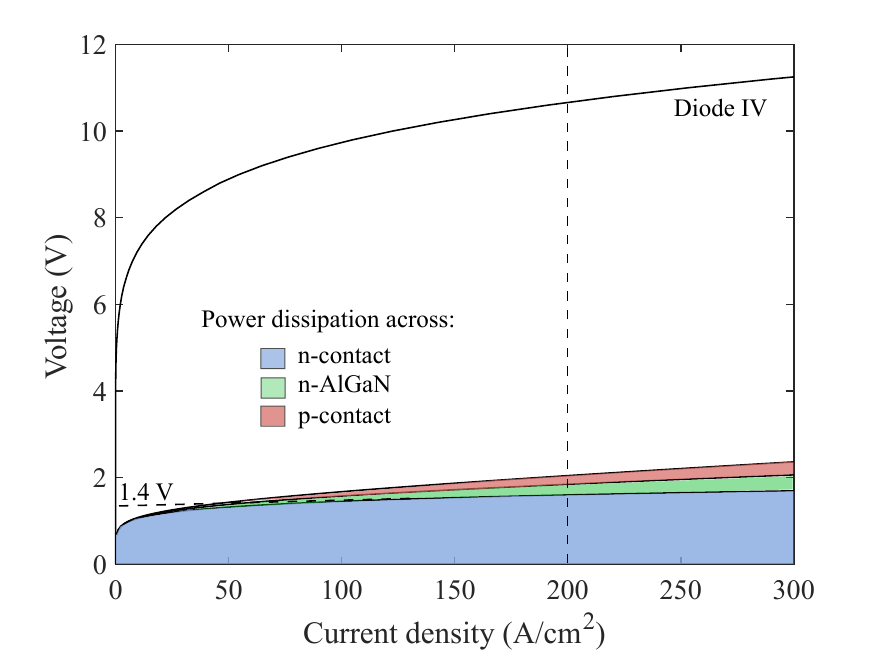}
    \caption{Voltage drop and power dissipation across different parasitic components as a function of LED current density.}
    \label{fig:Figure 4}
\end{figure}


\begin{figure*}[t]
   \includegraphics[width=0.8\textwidth]{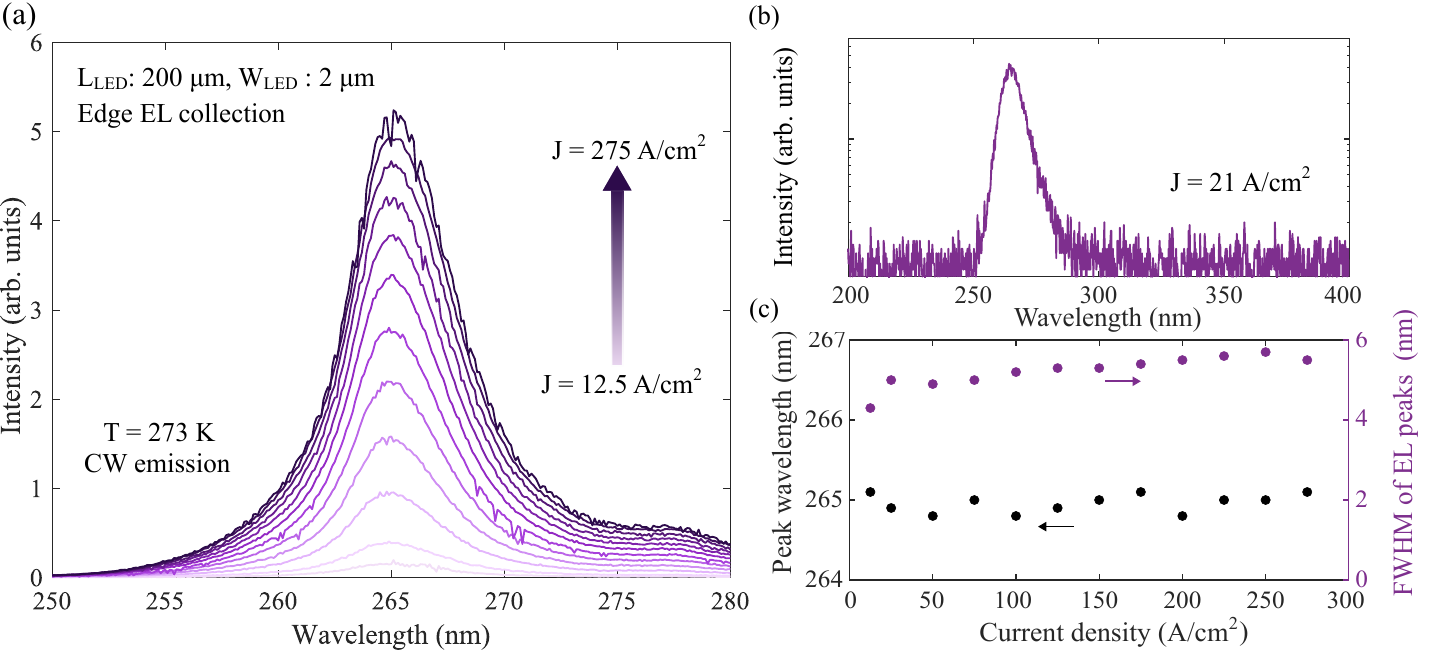}
    \caption{(a) Current dependent electroluminescence collected from a cleaved edge of the LED. (b) Electroluminescence in logarithmic scale in a wider range from 200-400 nm. (c) Dependence of the peak wavelength and the FWHM of the EL peaks on injection current.}
    \label{fig:Figure 6}
\end{figure*}
The $J$-$V$ characteristics of the diode at room temperature is shown in Fig.~\ref{fig:Figure 3}(d). An on/off ratio of $\sim$5-6 orders at $\pm$~5~V is observed. The LED turn-on voltage obtained from a linear extrapolation from a current density of 10~A/cm$^2$ is ~$\sim7.4$~V. The high turn-on voltage of the LED is from the non-ohmic metal/semiconductor contacts and is discussed more in the next section. 
 For the quasi-vertical device, we define the diode area as the p-electrode area. We use this area to report the room temperature differential on-resistance of the device evaluated as $R_\mathrm{on}\sim2$~m$\Omega\cdot$cm$^2$ at 800~A/cm$^2$ and 13~V.  Despite the sub-optimal n-contacts, the DUV LEDs were capable of achieving high injection current density on select devices, such as the one shown in Fig. \ref{fig:Figure 3}~(d). 



In Fig.~\ref{fig:Figure 4} we estimate the voltage drop across different parasitic resistances of the diode shown in Fig. \ref{fig:Figure 3}(d). The n-contact voltage drop is calculated as $V_{\mathrm{cn}}$~=~$\rho$$_{\mathrm{cn}}$~$\times$~$J_\mathrm{n}$, where the specific contact resistivity $\rho$$_{\mathrm{cn}}$ is extracted from the TLM data shown in Fig. \ref{fig:Figure 3}, and $J_\text{n}$ is the current density through the n-electrode. The \mbox{p-contact} voltage drop is calculated using the same method. The lateral resistance of the \mbox{n-AlGaN} layer is evaluated as $V_{\text{lateral}}$~=~$I$~$\times$~$R_{\text{n-AlGaN}}$, where $I$ is the current through the diode, and $$R_{\text{n-AlGaN}}= R_{\text{sh}}\times \frac{d}{L_{\text{LED}}}+R_{\text{sh}}\times \frac{W_{\text{LED}}}{2\times L_{\text{LED}}},$$ where $R_{\text{sh}}$ is the extracted sheet resistance, $d=17$~\textmu m is the lateral spacing between the n and p electrodes. The factor of 2 in the second term arises from the assumption that, under low-current conditions, the lateral current distribution decreases linearly from its maximum near the n-electrode mesa to zero at the opposite edge of the p-electrode. 

We identify most of the parasitic power to be dissipated across the n-contact, which introduces a turn-on voltage of $\sim$1.4~V in the diode, as extrapolated from the linear regime of the n-contact IV. This can be mitigated by lowering the n-contact resistance through higher Si doping\cite{bharadwaj2019bandgap} in n-Al$_{0.76}$Ga$_{0.24}$N and improving etching and annealing conditions. As shown in the dashed vertical line in Fig.~\ref{fig:Figure 4}, at a current density of 200~A/cm$^2$ and $\sim$10~V in the diode IV, about $\sim$2~V is accounted for in the n-contact, 0.21~V from the p-contact and 0.25~V from the n-AlGaN lateral resistance. About $\sim$5~V is needed for the quasi-Fermi level splitting in the active region, which leaves 2.5~V unaccounted for. The additional series resistance and delayed turn-on voltage may arise from heterojunction barriers (e.g., p-GaN/p-Al$_{0.65}$GaN$_{0.35}$, DPD-AlN/UID AlN) and from the UID 70~nm AlN interlayer and waveguides. AlGaN-based DUV LED processing studies have reported that the high temperature n-contact annealing degrades the p-contact\cite{Huang2025,hao2017improved,Bhattacharya_2025}.The power dissipation in the n-AlGaN lateral injection resistance can be reduced by minimizing the distance between the n and p-electrodes. While this approach is acceptable for LEDs, placing the p-electrode closer than $\sim$15~µm with respect to the mesa edge can increase the threshold current density of a laser diode \cite{kushimoto2021impact}, due to non-radiative recombination at the mesa edge from point defects caused by dry-etching. Increasing the Si doping concentration in the n-AlGaN can help reduce the lateral resistance.  


The EL spectrum is collected using an Acton SP2500 monochromator with a 240~nm blazed 2400~g/mm grating. Figure~\ref{fig:Figure 6}(a) shows the EL spectra collected from the cleaved $m$-plane facet of a 200~µm x 2~µm LED operating in CW mode at $T$ = 273~K  with increasing current density. The highest intensity peak emission wavelength is 265~nm ($\sim$~4.7~eV), along with a side peak emission at around 277~nm ($\sim$~4.5~eV) occuring at higher current densities. We believe that these 2 peaks are from the UID-Al$_{0.66}$Ga$_{0.34}$N waveguide core and the QW layer, respectively, given their peak positions and energy difference. The unwanted recombination in the waveguide layers could be because of electrons that overflow the well and are injected into the p-side of the 70~nm waveguide layer.  This was also seen during the development of UV-B laser diodes \cite{sato2020analysis}.  
Variants of the current heterostructure design with varying waveguide layer thicknesses and compositions will be studied in the future to confirm this hypothesis.  No parasitic luminescence is observed till 400~nm as shown in Fig.~\ref{fig:Figure 6}(b). The peak wavelengths do not shift with increasing current up to 275~A/cm$^2$. The low full width at half maximum (FWHM) of EL peaks of 5-6~nm indicate composition uniformity in the AlGaN layers.  A slight increase of FWHM with current can also be seen due to band filling in Fig.~\ref{fig:Figure 6}(c). 


To summarize, we have demonstrated 265~nm UV-C LEDs grown pseudomorphically on bulk AlN substrates by MBE. A high current density up to 800~A/cm$^2$ and on/off ratio of 5 orders of magnitude is achieved. We speculate that a high parasitic power dissipation occurrs at the n-contact. Therefore, to reduce operating voltages and achieve higher efficiency it is important to lower contact resistances to high composition n-AlGaN. Edge collection of photons was facilitated by a waveguide design and cleaved $m$-plane facets on the devices which are important for the development of UV-C lasers. High energy photons of 4.7~eV energy are collected from the ultrawide bandgap AlGaN active region. This device design and operation is important towards the development of UV-C optoelectronics on the AlN platform.

\section*{acknowledgments}
This work was supported by the Army Research Office
under Grant No. W911NF2220177 (characterization); ULTRA, an
Energy Frontier Research Center funded by the U.S. Department of Energy (DOE); SUPREME (modeling), one of seven centers in JUMP 2.0, a Semiconductor Research Corporation (SRC) program sponsored by DARPA; and the DARPA UWBGS program. This work was performed in part at the Cornell NanoScale Facility, an NNCI member supported by NSF Grant NNCI 2025233.
\vspace{-15pt}
\section*{author declarations}
S. Agrawal and H.W. Huang contributed equally to this work. 
\vspace{-15pt}
\section*{Conflict of Interest}
The authors have no conflict of interest to declare.
\section*{Data Availability Statement}
The data that support the findings of this study are available from the corresponding authors upon reasonable request.
\nocite{*}
\bibliography{citations}

@article{kneissl2019emergence,
  title={The emergence and prospects of deep-ultraviolet light-emitting diode technologies},
  author={Kneissl, Michael and Seong, Tae-Yeon and Han, Jung and Amano, Hiroshi},
  journal={Nature Photonics},
  volume={13},
  number={4},
  pages={233--244},
  year={2019},
  publisher={Nature Publishing Group UK London},
  url = {https://www.nature.com/articles/s41566-019-0359-9}
}

@article{zhang2024study,
  title={Study on Degradation of Deep-Ultraviolet Laser Diode},
  author={Zhang, Ziyi and Yoshikawa, Akira and Kushimoto, Maki and Sasaoka, Chiaki and Amano, Hiroshi},
  journal={Physica Status Solidi (a)},
  volume={221},
  number={21},
  pages={2300946},
  year={2024},
  publisher={Wiley Online Library},
  url = {https://onlinelibrary.wiley.com/doi/epdf/10.1002/pssa.202300946}

}

@article{agrawal2024ultrawide,
  title={Ultrawide bandgap semiconductor heterojunction p--n diodes with distributed polarization-doped p-type {AlGaN} layers on bulk {AlN} substrates},
  author={Agrawal, Shivali and van Deurzen, Len and Encomendero, Jimy and Dill, Joseph E and Wei Sheena Huang, Hsin and Protasenko, Vladimir and Xing, Huili Grace and Jena, Debdeep},
  journal={Applied Physics Letters},
  volume={124},
  number={10},
  year={2024},
  pages={102109},
  publisher={AIP Publishing},
  url={https://doi.org/10.1063/5.0189419}
}

@misc{SiLENSe,
  title = {{{SiLENSe, STR Software for Modeling of Crystal Growth, Epitaxy, and Semiconductor Devices}}},
  journal = {STR Software for Modeling of Crystal Growth, Epitaxy, and Semiconductor Devices},
  urldate = {2023-10-02},
  langid = {american},
  url={https://str-soft.com/devices/silense/}
}

@article{ramesh2025ultrawide,
  title={Ultrawide Bandgap Channel Polarization-Doped Junction Field-Effect Transistor},
  author={Ramesh, Madhav and Agrawal, Shivali and Xing, Huili Grace and Jena, Debdeep},
  journal={Physica Status Solidi (a)},
  pages={2401023},
  year={2025},
  publisher={Wiley Online Library},
  url={https://doi.org/10.1002/pssa.202401023}
}

@article{sato2020analysis,
  title={Analysis of Spontaneous Subpeak Emission from the Guide Layers of the {Ultraviolet-B} Laser Diode Structure Containing Composition-Graded {p-AlGaN} Cladding Layers},
  author={Sato, Kosuke and Yasue, Shinji and Ogino, Yuya and Iwaya, Motoaki and Takeuchi, Tetsuya and Kamiyama, Satoshi and Akasaki, Isamu},
  journal={Physica Status Solidi (a)},
  volume={217},
  number={14},
  pages={1900864},
  year={2020},
  publisher={Wiley Online Library},
  url={https://doi.org/10.1002/pssa.201900864}
}

@INPROCEEDINGS{9046469,
  author={Lee, Kevin and Bharadwaj, Shyam and Protasenko, Vladimir and Xing, Huili and Jena, Debdeep},
  booktitle={2019 Device Research Conference (DRC)}, 
  title={Efficient {InGaN} p-Contacts for deep-{UV} Light Emitting Diodes}, 
  year={2019},
  volume={},
  number={},
  pages={171-172},
  doi={10.1109/DRC46940.2019.9046469}}

@article{cho-2020,
	author = {Cho, YongJin and Chang, Celesta S. and Lee, Kevin and Gong, Mingli and Nomoto, Kazuki and Toita, Masato and Schowalter, Leo J. and Muller, David A. and Jena, Debdeep and Xing, Huili Grace},
	journal = {Applied Physics Letters},
	month = {4},
	number = {17},
	title = {{Molecular beam homoepitaxy on bulk AlN enabled by aluminum-assisted surface cleaning}},
	volume = {116},
	year = {2020},
    pages ={172106},
	url = {https://doi.org/10.1063/1.5143968},
}

@article{kumabe2024demonstration,
  title={Demonstration of {AlGaN-on-AlN} pn diodes with dopant-free distributed polarization doping},
  author={Kumabe, Takeru and Yoshikawa, Akira and Kawasaki, Seiya and Kushimoto, Maki and Honda, Yoshio and Arai, Manabu and Suda, Jun and Amano, Hiroshi},
  journal={IEEE Transactions on Electron Devices},
  volume={71},
  number={5},
  pages={3396--3402},
  year={2024},
  publisher={IEEE},
  url={https://ieeexplore.ieee.org/abstract/document/10445107}
}

@article{van2022optically,
  title={Optically pumped deep-{UV} multimode lasing in {AlGaN} double heterostructure grown by molecular beam homoepitaxy},
  author={van Deurzen, Len and Page, Ryan and Protasenko, Vladimir and Nomoto, Kazuki and Xing, Huili Grace and Jena, Debdeep},
  journal={AIP Advances},
  volume={12},
  number={3},
  year={2022},
  pages={035023},
  publisher={AIP Publishing},url = {https://doi.org/10.1063/5.0085365}
}

@article{kobayashi2024enhanced,
  title={Enhanced Wall-Plug Efficiency over 2.4\% and Wavelength Dependence of Electrical Properties at Far {UV-C} Light-Emitting Diodes on Single-Crystal {AlN} Substrate},
  author={Kobayashi, Hirotsugu and Sato, Kosuke and Okuaki, Yusuke and Lee, TaeGi and Kunimi, Yoshihisa and Kuze, Naohiro},
  journal={Physica Status Solidi (RRL)--Rapid Research Letters},
  volume={18},
  number={11},
  pages={2400002},
  year={2024},
  publisher={Wiley Online Library},
url = { https://doi.org/10.1002/pssr.202400002}
}

@article{zhang2022key,
  title={{Key temperature-dependent characteristics of AlGaN-based UV-C} laser diode and demonstration of room-temperature continuous-wave lasing},
  author={Zhang, Ziyi and Kushimoto, Maki and Yoshikawa, Akira and Aoto, Koji and Sasaoka, Chiaki and Schowalter, Leo J and Amano, Hiroshi},
  journal={Applied Physics Letters},
  volume={121},
  number={22},
  year={2022},
  pages={222103},
  publisher={AIP Publishing},url = {https://doi.org/10.1063/5.0124480}
}

@article{bagheri2022doping,
  title={Doping and compensation in heavily {Mg doped Al-rich AlGaN} films},
  author={Bagheri, Pegah and Klump, Andrew and Washiyama, Shun and Hayden Breckenridge, M and Kim, Ji Hyun and Guan, Yan and Khachariya, Dolar and Qui{\~n}ones-Garc{\'\i}a, Cristyan and Sarkar, Biplab and Rathkanthiwar, Shashwat and others},
  journal={Applied Physics Letters},
  volume={120},
  number={8},
  year={2022},
  pages={082102},
  publisher={AIP Publishing},
url = {https://doi.org/10.1063/5.0082992}
}

@article{ruschel2020reliability,
  title={Reliability of {UVC LEDs fabricated on AlN/sapphire} templates with different threading dislocation densities},
  author={Ruschel, Jan and Glaab, Johannes and Susilo, Norman and Hagedorn, Sylvia and Walde, Sebastian and Ziffer, Eviathar and Cho, Hyun Kyong and Ploch, Neysha Lobo and Wernicke, Tim and Weyers, Markus and others},
  journal={Applied Physics Letters},
  volume={117},
  number={24},
  year={2020},
  pages={241104},
  publisher={AIP Publishing},
url = {https://doi.org/10.1063/5.0027769}
}

@article{zhang2021polarization,
  title={Polarization-induced {2D hole gases in pseudomorphic undoped GaN/AlN heterostructures on single-crystal AlN substrates}},
  author={Zhang, Zexuan and Encomendero, Jimy and Chaudhuri, Reet and Cho, Yongjin and Protasenko, Vladimir and Nomoto, Kazuki and Lee, Kevin and Toita, Masato and Xing, Huili Grace and Jena, Debdeep},
  journal={Applied Physics Letters},
  volume={119},
  number={16},
  year={2021},
  pages={162104},
  publisher={AIP Publishing},
url = {https://doi.org/10.1063/5.0066072}
}

@article{ren2007algan,
  title={{AlGaN deep ultraviolet LEDs on bulk AlN} substrates},
  author={Ren, Zaiyuan and Sun, Q and Kwon, S-Y and Han, J and Davitt, K and Song, YK and Nurmikko, AV and Liu, W and Smart, J and Schowalter, L},
  journal={Physica Status Solidi C},
  volume={4},
  number={7},
  pages={2482--2485},
  year={2007},
  publisher={Wiley Online Library},
url = { https://doi.org/10.1002/pssc.200674758}
}

@article{chen2021gan,
  title={Ga{N} buffer growth temperature and efficiency of {InGaN/GaN} quantum wells: The critical role of nitrogen vacancies at the {GaN} surface},
  author={Chen, Yao and Haller, Camille and Liu, Wei and Karpov, Sergey Yu and Carlin, Jean-Fran{\c{c}}ois and Grandjean, Nicolas},
  journal={Applied Physics Letters},
  volume={118},
  number={11},
  pages={111102},
  year={2021},
  publisher={AIP Publishing},
  url={https://doi.org/10.1063/5.0040326}
}

@article{kushimoto2021impact,
  title={Impact of heat treatment process on threshold current density in {AlGaN}-based deep-ultraviolet laser diodes on {AlN} substrate},
  author={Kushimoto, Maki and Zhang, Ziyi and Sugiyama, Naoharu and Honda, Yoshio and Schowalter, Leo J and Sasaoka, Chiaki and Amano, Hiroshi},
  journal={Applied Physics Express},
  volume={14},
  number={5},
  pages={051003},
  year={2021},
  publisher={IOP publishing},doi = {10.35848/1882-0786/abf443}
}

@article{hao2017improved,
  title={Improved turn-on and operating voltages in {AlGaN}-based deep-ultraviolet light-emitting diodes},
  author={Hao, Guo-Dong and Taniguchi, Manabu and Tamari, Naoki and Inoue, Shin-ichiro},
  journal={Journal of Electronic Materials},
  volume={46},
  pages={5677--5683},
  year={2017},
  publisher={Springer},
doi = {10.1007/s11664-017-5622-6}
}

@article{bharadwaj2019bandgap,
  title={Bandgap narrowing and {Mott transition in Si-doped Al$_{0.7}$Ga$_{0.3}$N}},
  author={Bharadwaj, Shyam and Islam, SM and Nomoto, Kazuki and Protasenko, Vladimir and Chaney, Alexander and Xing, Huili Grace and Jena, Debdeep},
  journal={Applied Physics Letters},
  volume={114},
  number={11},
  year={2019},
  pages={113501},
  publisher={AIP Publishing},url = {https://doi.org/10.1063/1.5086052}
}

@article{hoke2007thermodynamic,
  title={Thermodynamic analysis of cation incorporation during molecular beam epitaxy of nitride films using metal-rich growth conditions},
  author={Hoke, WE and Torabi, A and Mosca, JJ and Kennedy, TD},
  journal={Journal of Vacuum Science \& Technology B: Microelectronics and Nanometer Structures Processing, Measurement, and Phenomena},
  volume={25},
  number={3},
  pages={978--982},
  year={2007},
  publisher={AIP Publishing},
  url={https://doi.org/10.1116/1.2716003}
}

@article{Huang2025,
  title={Low p-contact resistance {InGaN-capped AlGaN-based DUV LEDs on bulk AlN substrates}},
  author={Huang, Hsin-Wei S and Agrawal, Shivali and Bhattacharya, Debaditya and Protasenko, Vladimir and Xing, Huili Grace and Jena, Debdeep},
  journal={Applied Physics Letters},
  volume={127},
  number={19},
  year={2025},
pages={193305},
  publisher={AIP Publishing},  doi={10.1063/5.0297626}
}

@unpublished{Bhattacharya_2025,
    author = {Bhattacharya, Debaditya and Agrawal, Shivali and Huang, Hsin-Wei S. and Ramesh, Madhav and Protasenko, Vladimir and Xing, Huili Grace and Jena, Debdeep}, 
    title = {{Dielectric-assisted liftoff enabled simultaneous low n- and p-contact resistivities in ultrawide bandgap {AlGaN} pn diodes on bulk AlN}},
    journal = {Japanese Journal of Applied Physics},
    year = {forthcoming in 2026}
}

\end{document}